# Charge transfer induced insulating state at antiperovskite/perovskite heterointerfaces

**Authors:** Ting Cui,[1,2,†] Ying Zhou,[3,†] Qianying Wang,[1,2] Dongke Rong,[1,2] Haitao Hong,[1,2] Axin Xie,[1,2] Jun-Jie Zhang,[3] Qinghua Zhang,[1] Can Wang,[1,2] Chen Ge,[1,2] Lin Gu,[4] Shanmin Wang,[5] Kuijuan Jin,[1,2,*] Shuai Dong,[3,*] and Er-Jia Guo[1,2,*]

**Affiliations:**

[1] *Beijing National Laboratory for Condensed Matter Physics and Institute of Physics, Chinese Academy of Sciences, Beijing 100190, China*

[2] *Department of Physics & Center of Materials Science and Optoelectronics Engineering, University of Chinese Academy of Sciences, Beijing 100049, China*

[3] *Key Laboratory of Quantum Materials and Devices of Ministry of Education, School of Physics, Southeast University, Nanjing 211189, China*

[4] *National Center for Electron Microscopy in Beijing and School of Materials Science and Engineering, Tsinghua University, Beijing 100084, China*

[5] *Department of Physics, Southern University of Science and Technology, Shenzhen 518055, China*

†These authors contribute equally to the manuscript.

*E-mails: kjjin@iphy.ac.cn, sdong@seu.edu.cn, and ejguo@iphy.ac.cn

**Abstract**

Heterointerfaces have been pivotal in unveiling extraordinary interfacial properties and enabling multifunctional material platforms. Despite extensive research on all-oxide interfaces, heterointerfaces between different material classes, such as oxides and nitrides, remain underexplored. Here we present the fabrication of a high-quality Dirac metal antiperovskite $Ni_3InN$, characterized by an extremely low temperature coefficient of resistivity, approximately $1.8 \times 10^{-8}$ Ω·cm/K, over a broad temperature range. Atomically sharp heterointerfaces between $Ni_3InN$ and $SrVO_3$ were constructed, revealing intriguing interfacial phenomena. Leveraging layer-resolved scanning transmission electron microscopy and electron energy loss spectroscopy, we identified pronounced charge transfer across the well-ordered interface. Remarkably, this interfacial electron transfer from $Ni_3InN$ to $SrVO_3$ induces an insulating interfacial layer and an emergent magnetic moment within the $Ni_3InN$ layer, consistent with first-principles calculations. These findings pave the way for novel electronic and spintronic applications by enabling tunable interfacial properties in nitride/oxide systems.

**Main text**

Heterointerfaces are a fundamental aspect of modern materials science, enabling the discovery of emergent phenomena arising from the interaction of distinct material properties. Two-dimensional electron gas (2DEGs) [1], polar metal [2], strong polarization enhancement [3], unexpected magnetic anisotropy [4] and exchange bias [5,6] had been discovered over decades. In the past, research has primarily focused on heterointerfaces constructed by semiconductors, all-oxides and metals, while the exploration of other innovative interface systems has been relatively limited. Advances in thin-film growth techniques now allow for the precise fabrication of high-crystallinity nitrides, facilitating the creation of nitride/oxide heterointerfaces [7-9].

Nitride/oxide interfaces represent an exciting frontier, combining the robustness and high-performance characteristics of nitrides with the functional diversity of oxides [8]. Except for the binary nitrides with simple crystal structure, antiperovskite nitrides are known for their exceptional electronic and thermal properties [10-16]. More importantly, the structures of antiperovskite nitrides resemble that of perovskites, with the cation and anion positions swapped, sharing the similar crystal structure and lattice constants as most of perovskite oxides, enabling the heterogeneous those two classes of materials at an atomic-scale interface [17]. Therefore, this type of heterointerfaces holds promise for discovering novel electronic, magnetic, and optical behaviors, particularly in the realm of quantum materials [13,18].

A key phenomenon in heterointerfaces is interfacial charge transfer, where electron redistribution occurs due to differences in electronic structure and chemical potential [19-24]. This phenomenon often gives rise to unexpected and emergent properties, including interfacial conductivity [25,26], magnetism [27], and topological states [28]. However, when charge transfer occurs in systems with strong electron correlations, as in the case of many oxides [29-32], the dynamics become highly complex. In nitride/oxide interfaces, these challenges are amplified by stark differences in bonding nature, orbital hybridization, and chemical environments. As a result, charge carrier migration across the interface can profoundly influence the overall functionality of these heterostructures.

Despite extensive research on all-oxide interfaces, nitride/oxide interfaces remain underexplored. This is particularly true for correlated metallic interfaces, where the coupling between itinerant electrons and local magnetic (or structural degrees of freedom) plays a critical role [8]. Interfacial charge transfer at these interfaces is both a challenge and an opportunity, offering insights into interfacial charge dynamics and emergent quantum phenomena. For instance, interfacial charge transfer can influence magnetic ordering, modify the density of states near the Fermi level, and create polarization effects that alter transport properties. Addressing these challenges requires an integrated approach, combining advanced synthesis techniques with atomic-scale characterization and theoretical modeling.

In this study, we report the synthesis of single-crystalline Dirac metal antiperovskite nitride thin films of nickel indium nitride (Ni₃InN), which exhibit an exceptionally low temperature coefficient of resistivity, making them ideal for investigating charge transport phenomena. Our focus is on the heterointerface between antiperovskite Ni₃InN and the correlated metallic perovskite oxide SrVO₃, characterized by a partially filled $d$-electron band and strong electron-electron interactions. By fabricating atomically sharp heterointerfaces and utilizing a combination of layer-resolved scanning transmission electron microscopy (STEM) and electron energy loss spectroscopy (EELS), we systematically examine the interfacial charge transfer mechanisms. Additionally, first-principles calculations are employed to elucidate the microscopic origin of interfacial charge transfer and its influence on the system's electronic and magnetic properties.

**Synthesis and characterization of antiperovskite nitride Ni₃InN thin films**

The electronic band structure and density of states (DOS) of bulk Ni₃InN were calculated using density functional theory (DFT), as shown in Figures 1a and 1b. According to the DOS, Ni's $3d$ orbitals dominate the electronic states around the Fermi surface, with a few hybridized contributions from anions. Interestingly, the band structure reveals that Ni₃InN is a Dirac metal with two Dirac nodes near the Fermi level, exhibiting a Dirac-like linear energy-momentum dispersion. These features contribute to the extremely high conductivity and high carrier mobility of Ni₃InN, consistent with the electrical properties presented in Figure 2, similar to that in graphene and topological semimetals. Furthermore, finite local magnetic moment at Ni site can be safely excluded by tuning the Hubbard $U$ in our DFT calculation (see Supplemental Materials), implying the non-magnetic nature of Ni₃InN. Therefore, the electronic configuration of Ni should be in the low spin state.

Ni₃InN thin films were fabricated on (001)-oriented (La₀.₃Sr₀.₇)(Al₀.₆₅Ta₀.₃₅)O₃ (LSAT) substrates using pulsed laser deposition (see Methods in Supplemental Materials). Ideal Ni₃InN adopts a cubic antiperovskite structure with space group $Pm\bar{3}m$, featuring an in-plane lattice parameter $a$ = 3.84 Å. The nitrogen atom resides at the center of octahedra, surrounded by Ni atoms at the face-centered positions and In atoms at the unit cell corners. XRD $\theta$-$2\theta$ scans confirmed the single-phase nature of Ni₃InN, as evidenced by the observation of only (00l) reflections (Figure S1). Phi scans and reciprocal space mapping (RSM) analysis revealed the fourfold crystal symmetry of Ni₃InN and cube-on-cube epitaxy on LSAT substrates. High-resolution cross-sectional high-angle annular dark field (HAADF) and integrated differential phase contrast (iDPC) imaging were conducted using scanning transmission electron microscopy (STEM) to verify the high crystallinity and cubic structure of Ni₃InN. Figures 1c and 1d present representative STEM images of antiperovskite Ni₃InN thin films along the [100] and [110] orientations, respectively. Atomic intensity in the images correlates with the atomic number ($Z$) of elements, and nitrogen atoms were distinctly resolved in the iDPC image along the [110] orientation. Structural analysis demonstrated that the as-grown Ni₃InN thin films exhibit excellent crystallinity and a well-ordered antiperovskite structure. X-ray absorption spectroscopy (XAS) was performed to investigate the valence states of transition metal ions in Ni$_3$InN thin films (Figure S2). Spectra collected at the N $K$- and Ni $L$-edges indicated a significant presence of nitrogen in the films. The majority of Ni ions were in the +3 oxidation state, while a minor fraction of Ni$^{2+}$ suggested slight surface oxidation of the Ni$_3$InN layers. Indeed, such an odd electron occupation per Ni not only leads to robust Dirac metallicity in the non-magnetic state but also allows for the possibility of Mottness and the emergence of a finite magnetic moment under suitable physical conditions.

**Nearly zero temperature coefficient of resistivity in Ni$_3$InN**

We measured the electrical transport behavior of the Ni$_3$InN thin films. Figure 2a shows the temperature-dependent resistivity ($\rho$) curves of a 30 nm-thick Ni$_3$InN thin film. The resistivity remains nearly constant across the entire temperature range, with a low-temperature resistivity upturn below 20 K. The $\rho$-$T$ curve of Ni$_3$InN thin films is fitted into two parts. The high-temperature regime (from 65 to 300 K) shows a linear temperature dependence, described by the formula $\rho(T) = \rho_1 + \beta T$, where $\rho_1$ is the residual resistivity due to defect scattering, and $\beta T$ represents the contribution from electron-phonon scattering, which dominates at higher temperatures. The parameter $\beta$ reflects the strength of the electron-phonon interaction, which is typically related to lattice vibrations and their coupling with conduction electrons. The low-temperature regime (from 5 to 65 K) can be well fitted by the formula $\rho(T)=\rho_0 + AT^2$, where $\rho_0$ is the residual resistivity due to defects or impurities, and $A$ represents the $T^2$-term coefficient. The fitted large $A$ indicates that Ni$_3$InN thin films exhibit Fermi liquid behavior with strong electron-electron scattering at low temperatures.

Figure 2b shows the carrier density ($n$) and Hall mobility ($\mu$) of Ni$_3$InN thin films as a function of temperature, calculated from Hall measurements. The $n$ of Ni$_3$InN remains stable in the range of $(7.81-8.81)\times 10^{22}$ cm$^{-3}$, and the $\mu$ exhibits minimal variation across the temperature range from 5 to 300 K. These behaviors are closely associated with the resistance stability characteristic of Ni$_3$InN thin films. To further investigate the electronic scattering mechanisms, we measured the magnetoresistance (MR) at different temperatures. As shown in Figure 2c, the magnetoresistance (defined as MR=$(R-R_0)/R_0$) is positive and increases with the magnetic field, with a more pronounced effect at lower temperatures. We attribute this behavior to the magnetic field suppressing both coherent hopping (band-like transport) and variable-range hopping (VRH). As a result, electron transport becomes increasingly difficult with a stronger magnetic field, leading to the observed positive MR. [36-38]

It is interesting that the value of resistivity at high temperature region is rather stable: d$\rho$/d$T$ is estimated as ~1.8×10$^{-8}$ Ω·cm/K over a wide temperature regime. Figure 2d and Table I summarizes the d$\rho$/d$T$ values of reported antiperovskite nitrides, manganin, and metals. Previous studies have demonstrated that Cu$_3$PdN possesses a relatively small temperature coefficient of resistivity (TCR) and has been proposed as a three-dimensional Dirac semimetal [10-12]. Our work shows that the TCR of Ni$_3$InN is even an order of magnitude lower than those of other antiperovskite nitrides [10,33-35]. The inset of Figure 2d illustrates the minimal $d\rho/dT$ value versus the linear temperature range ($\Delta T$). We defined that $\Delta T$ represents the temperature range over which linear resistance is maintained. The larger $\Delta T$ shows the ability of maintaining the low TCR for a material. Our work indicates that the resistance of Ni$_3$InN remains nearly unchanged over approximately 275 K regime. This stability far surpasses that of all previously reported antiperovskite nitrides. Additionally, we calculated the transport behavior of single-crystalline Ni$_3$InN (Figure S3). The calculated resistivity, variation trends, and low-temperature behavior are in excellent agreement with our experimental results qualitatively. Temperature-dependent *θ-2θ* scans of Ni$_3$InN single-crystalline thin films (Figure S4) reveal the structural variation as a function of temperature. There is no structural phase transition and only a slight change in the *c*-axis lattice constant from 10 to 300 K due to thermal expansion. We anticipate that the dual robustness of the crystalline structure and Dirac metallicity of Ni$_3$InN is the primary mechanism behind its remarkable resistance stability, an intrinsic property of Ni$_3$InN thin films.

**Interfacial charge transfer from Ni to V evidenced by layer-resolved EELS**

To investigate the intriguing interfacial charge transfer and its impact on transport behavior, we selected the antiperovskite nitride Ni$_3$InN and another correlated metallic SrVO$_3$ as a model system. This choice was motivated by the multiple valence states of Ni and V, as well as their distinct Fermi levels. Their structural compatibility and unique electronic properties offer an exciting platform to explore interfacial charge transfer. We fabricated Ni$_3$InN/SrVO$_3$ heterostructures using PLD. The high-quality coherent epitaxial growth was confirmed through XRD *θ-2θ* scans and reciprocal space mappings (RSMs) (Figure S5). To examine the microscopic structures and chemical compositions, we performed atomic-resolution HADDF-STEM imaging and EELS (Figure 3a and S6). The heterointerfaces revealed atomically sharp boundaries with negligible chemical intermixing. EELS analysis further confirmed the presence of a clear nitrogen signal in the Ni$_3$InN thin films, indicating correct stoichiometry and high crystallinity. These structural and compositional insights provide a robust foundation for deeper exploration of interfacial phenomena and their potential implications for advanced electronic systems.

To exploit the interfacial charge transfer across the Ni$_3$InN/SrVO$_3$ interface, we conducted a series of EELS line scans from the Ni$_3$InN layer to the SrVO$_3$ layer to investigate the layer-resolved electronic structure. The EELS spectra were acquired at both the Ni $L_{2,3}$- and V $L_{2,3}$-edges across the heterointerface. Figures 3b and 3c illustrate the evolution of the electronic structures of Ni and V, respectively. In addition to variations in spectral intensity, significant peak shifts were observed: the Ni $L_{2,3}$ peaks increased by ~0.6 eV from the Ni$_3$InN layer to the interface, whereas the V $L_{2,3}$ peaks decreased by ~0.8 eV from the SrVO$_3$ layer to the interface. Detailed comparisons of these peak shifts are presented in Figure 3d. These pronounced differences between the SrVO$_3$ and Ni$_3$InN layers reveal a clear interfacial charge transfer, with electrons transferring from the Ni$_3$InN layer to the SrVO$_3$ layer. The most significant peak shifts were observed near the interface region, with the extent of charge transfer diminishing gradually as the layers extended further from the interface. The charge-modified region was confined to a thickness of 2–3 unit cells, indicating that the charge transfer is highly localized.

DFT calculations were conducted to validate charge transfer at the Ni$_3$InN/SrVO$_3$ interface and analyze its electronic structure. First, we calculated the electronic and magnetic properties of SrVO$_3$ (details in Supplemental Material). The V$^{4+}$ ion exhibits a finite magnetic moment of 1 µ$_B$/V, consistent with its 3$d^1$ configuration. However, the energy difference between the ferromagnetic state and A-type antiferromagnetic state is very small, suggesting a low magnetic ordering temperature, if not zero, which aligns with the paramagnetic nature of SrVO$_3$ observed in our experiments. Next, we evaluated the formation energy (ΔE) of five possible interfacial configurations between Ni$_3$InN and SrVO$_3$ (Figure S7). Among these, the Ni$_2$N-VO$_2$ termination exhibited the lowest ΔE, indicating it as the most stable interface configuration (Figure 4a). This theoretical prediction is in excellent agreement with our atomic-resolution HAADF-STEM imaging (Figure 4b and S8). To further confirm interfacial charge transfer, we performed charge density differential calculations and Bader charge analysis across the Ni$_3$InN/SrVO$_3$ interface (Figures 4c, 4d, and S9). The results reveal substantial electron redistribution localized near the interface, with minimal changes in the bulk regions of both materials. Specifically, Ni atoms at the first interfacial layer exhibit electron depletion (–0.73e$^-$). Due to the broad spatial distribution of electron clouds in this metallic system, our DFT calculations provide semiquantitative rather than strictly quantitative values. Nonetheless, these results unambiguously confirm electron transfer from Ni$_3$InN to SrVO$_3$, in excellent agreement with our experimental observations.

As previously discussed, Ni ions in Ni$_3$InN adopt a low-spin state; however, magnetism can emerge under specific physical conditions. To further investigate the impact of interfacial coupling between the Ni$_3$InN and SrVO$_3$ layers, we examined the evolution of magnetic moments across atomic layers (Figure 4e). Remarkably, a strong local magnetic moment of ~0.6 µ$_B$/Ni emerges at the first interfacial Ni layer, despite the nominally nonmagnetic nature of bulk Ni$_3$InN. Concurrently, the local magnetic moment of V in SrVO$_3$ is enhanced from ~1 µ$_B$/V to ~1.5 µ$_B$/V at the first interfacial V layer. These interfacial moments exhibit an antiferromagnetic alignment between Ni and V, while the first two interfacial V layers become ferromagnetically coupled.

The emergence of magnetism at the Ni$_3$InN/SrVO$_3$ interface is directly linked to the unconventional charge transfer across the heterostructure. In a typical oxide interface, charge transfer is primarily driven by band alignment and electrostatic potential differences. However, in this case, the presence of a Dirac-like metallic Ni$_3$InN layer with low-spin Ni ions creates an environment where charge redistribution can induce unexpected magnetic behavior. This unique behavior can be concluded into three aspects. Firstly, the electronic configuration of Ni has been modified by bonding enviroment. In bulk Ni$_3$InN, Ni is in a low-spin state with a nominally nonmagnetic character due to strong hybridization between Ni 3*d* and N 2*p* orbitals. However, at the interface, electron depletion in the Ni layer (~ –0.73e$^-$ per Ni$_2$N layer) disrupts this hybridization, shifting the Ni 3*d* orbitals closer to an intermediate-spin or high-spin state. This results in the emergence of a localized magnetic moment at interfacial Ni sites, which is absent in bulk Ni$_3$InN. Secondly, the V magnetic moment in SrVO$_3$ enhances due to the charge transfer. The electron accumulation at interfacial V sites (~ +0.13e$^-$ per VO$_2$ layer) increases orbital occupancy and Coulomb interactions, reinforcing magnetic exchange interactions. The increased local moment suggests an interfacial-driven magnetic phase transition. Lastly, the induced Ni moments and the enhanced V moments are antiferromagnetically coupled at the interface. This behavior resembles superexchange interactions in correlated electron systems, where charge transfer modulates magnetic ordering. Additionally, the first two V layers exhibit ferromagnetic coupling, likely due to interface-mediated double exchange or kinetic energy stabilization.

In conclusion, we have successfully fabricated high-quality single-crystalline Ni$_3$InN thin films and epitaxial heterostructures composed of antiperovskite nitride Ni$_3$InN and perovskite oxide SrVO$_3$ (SVO). The Ni$_3$InN thin films demonstrate nearly constant electrical resistance over a wide temperature range, making them highly suitable for applications such as precision resistors, stable sensors, and other high-precision instruments operating across diverse temperature environments. Furthermore, interfacial charge transfer near the Ni$_3$InN/SVO interface was confirmed through layer-resolved EELS and first-principles calculations. Unexpectedly, our findings reveal that charge transfer from Ni$_3$InN to SrVO$_3$ induces an unexpected magnetic moment within the Ni$_3$InN layer. These findings offer valuable insights into interfacial charge transfer mechanisms at nitride/oxide interfaces, highlighting their potential for next-generation electronic, magnetic, and energy applications.

**Table and table captions**

| Materials | TCR (ppm/K) | $\Delta T$ (K) | Ref. |
|---|---|---|---|
| Manganin | / | 150 | [33] |
| $Mn_3AgN$ | 47 | 65 | [34] |
| $Mn_3CuN$ | 46 | 150 | [33] |
| $Mn_3CoN$ | 310 | 129 | [35] |
| $Mn_3NiN$ | 540 | 141 | [35] |
| $Mn_3ZnN$ | 810 | 200 | [35] |
| $Mn_3PdN$ | 1090 | 31 | [35] |
| $Cu_3PdN$ | 150 | 233 | [10] |
| Cu | 4928 | 190 | [39] |
| Al | 529 | 100 | [40] |
| Au | 2173 | 275 | [41] |
| Ag | 2200 | 275 | [41] |
| This is work | 167 | 275 | / |

**Table I.** Summary of TCR values for previously reported antiperovskite nitrides and for $Ni_3InN$ presented in this work. The metals were listed for comparison. ΔT represents the temperature range over which linear resistance is maintained.

**Figures and figure captions**

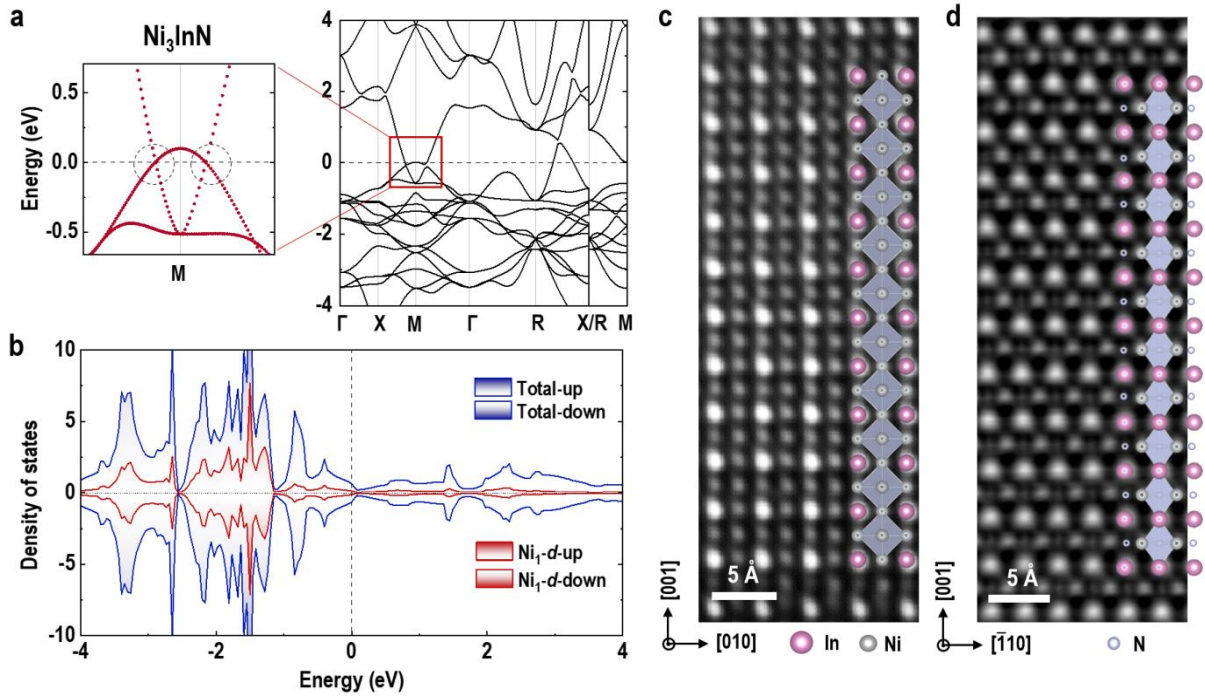

**Figure 1. Band structure calculations and structural characterization of Ni₃InN.**
(a) Band structure of Ni₃InN, with the red square highlighting the detailed band structure near the *M* point. Two distinct nodes are evident along the *X-M* and *M-Γ* momentum directions. (b) Total and projected density of states (DOS) for Ni₃InN, showing near symmetry between spin-up and spin-down electrons. The non-zero DOS at the Fermi level confirms that Ni₃InN is a non-magnetic Dirac metal. (c) High-resolution HAADF-STEM image and (d) iDPC-STEM image of a Ni₃InN film, viewed along the [100] and [110] orientations, respectively. These images validate the anti-perovskite structure composed of N atoms.

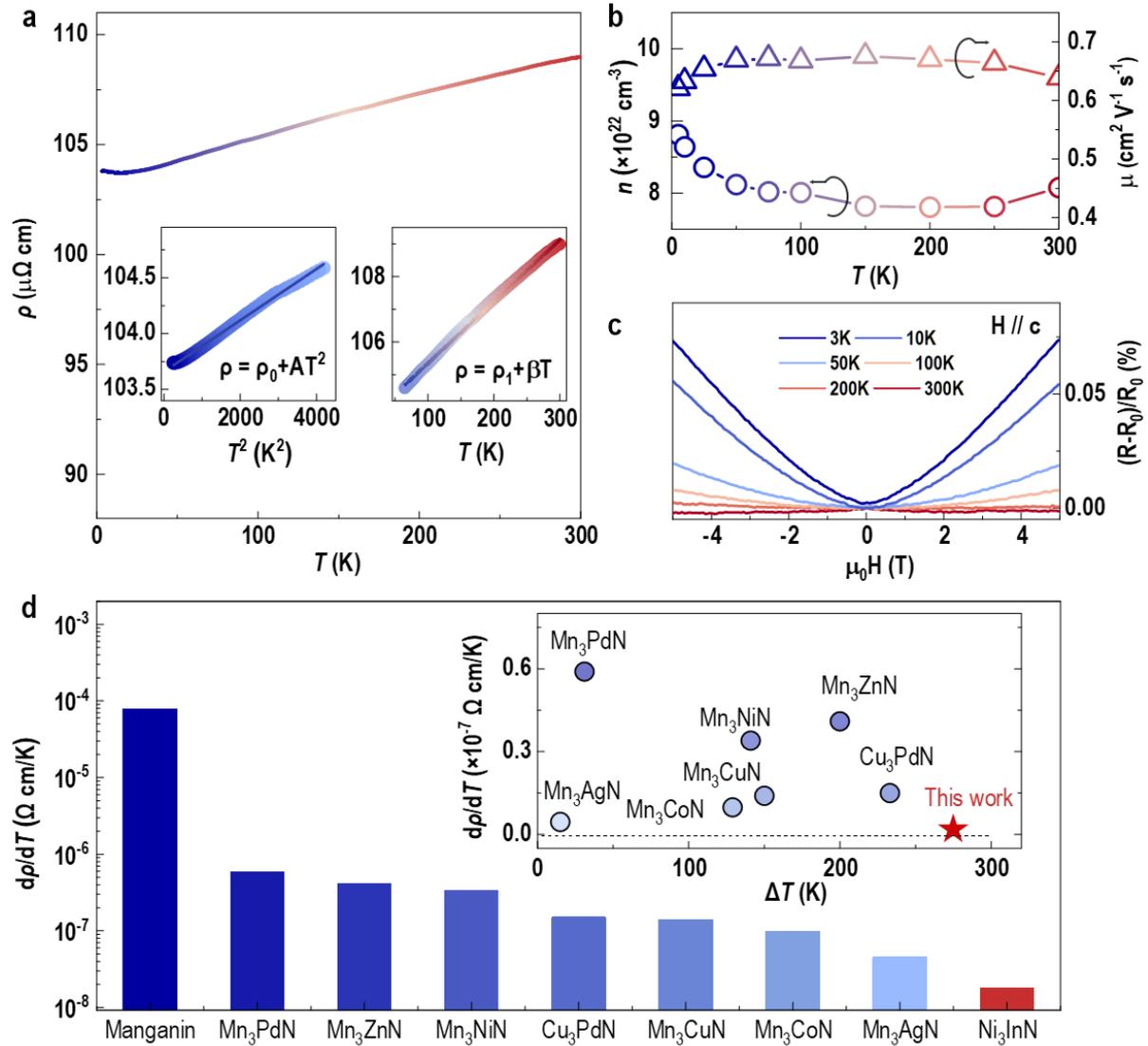

**Figure 2. Ultralow temperature coefficient of resistance (TCR) in Ni₃InN thin films.**
(a) Temperature-dependent resistivity (ρ) of Ni₃InN. The insets illustrate the linear fits to different models. (b) Carrier densities and Hall mobilities of Ni₃InN as functions of temperature. (c) Field-dependent magnetoresistance measured at various temperatures. (d) Comparison of d$\rho$/d$T$ values of Ni₃InN with other antiperovskite nitrides. The inset of (d) summarizes d$\rho$/d$T$ for antiperovskite nitrides across a wide temperature range.

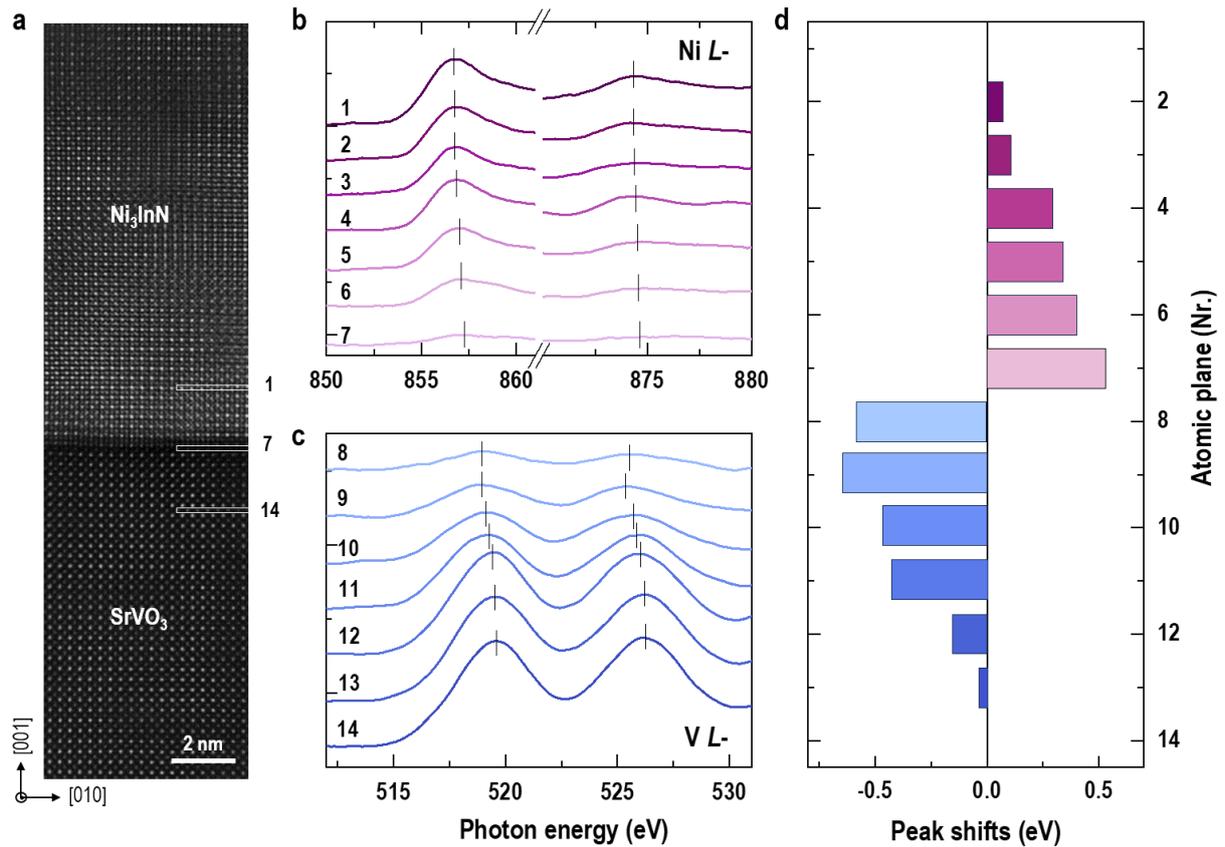

**Figure 3. Interfacial charge transfer at the Ni₃InN/SrVO₃ Interface.**
(a) HAADF-STEM image of the Ni₃InN/SrVO₃ interface, with an EELS line scan across the interface. (b) and (c) EELS spectra at the Ni $L$- and V $L$-edge, respectively. The numbers in (a) represent the atomic layers across the interface. (d) Peak shifts of the EELS spectra as a function of atomic layers. The Ni $L$-edge peaks shift to higher energies, while the V $L$-edge peaks shift to lower energies moving from the bulk to the interface, indicating an increase in the valence states of Ni and a decrease in the valence states of V.

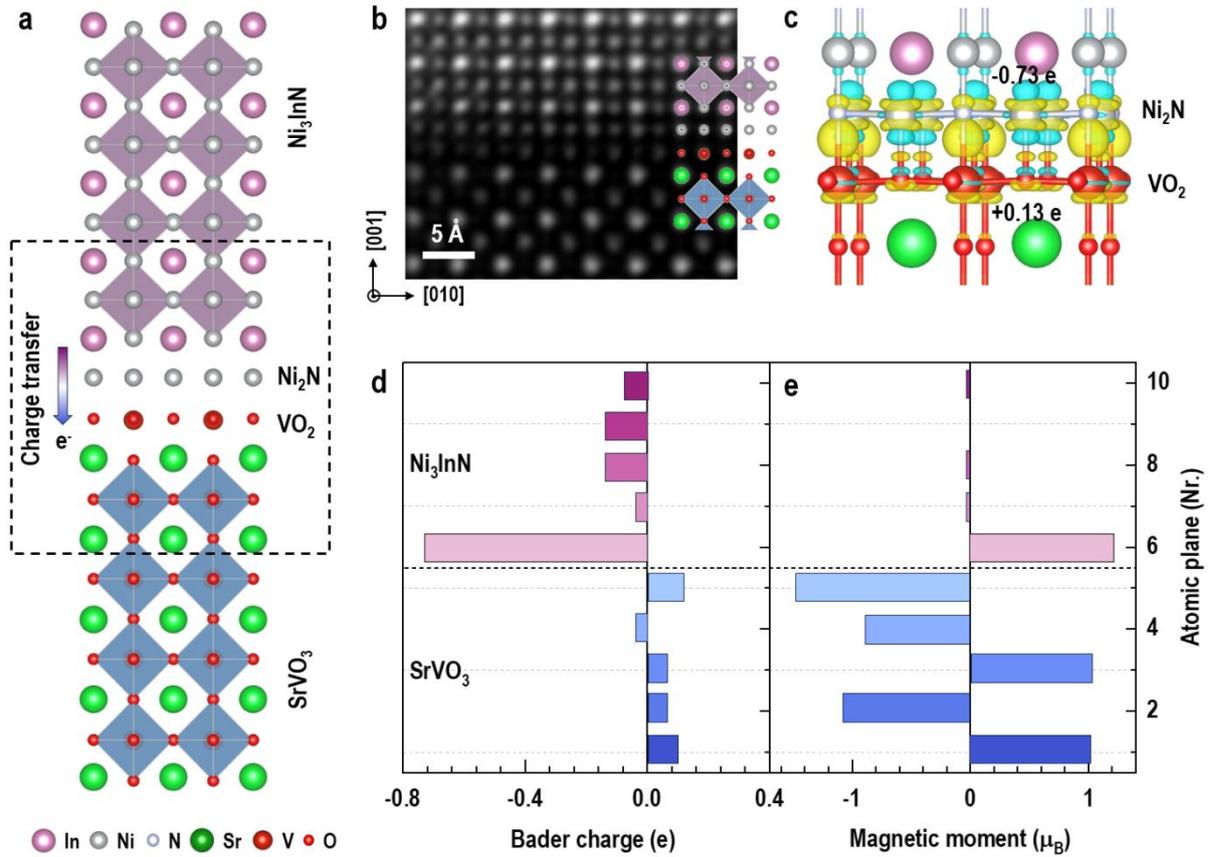

**Figure 4. First-principles calculations of the interfacial charge states and magnetism at the Ni₃InN/SrVO₃ interfaces.**

(a) Schematic illustration of the lowest-energy Ni₃InN/SrVO₃ interface with Ni₂N-VO₂ termination. (b) High-resolution HAADF-STEM image of the Ni₃InN/SrVO₃ heterostructures, confirming the Ni₂N-VO₂ interfacial configuration. (c) Electronic density difference distributions at the interface, with isosurfaces indicating electron accumulation (yellow) and charge depletion (blue), comparing with the isolated Ni₃InN and SrVO₃ slabs. (d) Layer-resolved Bader charge and (e) magnetic moment profiles across the interface.